\documentclass[prl,twocolumn,twoside,showpacs,floatfix,superscriptaddress]{revtex4}

\usepackage{graphicx,color,rotating}
\usepackage{amsmath,amssymb}
\usepackage{dcolumn}
\usepackage{times,txfonts,pifont}

\newcommand{\ket}[1]{\ensuremath{\,|{#1}\rangle}}

\newcommand{\comm}[2]{\ensuremath{[{#1},{#2}]}}

\newcommand{\op}[1]{\ensuremath{#1}}

\newcommand{\elem}[2]{\ensuremath{{}^{#2}\text{#1}}}

\newcommand{\symboldiamond}[1][black]{{\color{#1}\hspace{-1pt}\footnotesize\begin{turn}{45} $\blacksquare$ \end{turn}}}
\newcommand{\symboltriangle}[1][black]{{\color{#1}$\blacktriangle$}}
\newcommand{\symbolbox}[1][black]{{\color{#1}$\blacksquare$}}
\newcommand{\symbolcircle}[1][black]{{\color{#1}\large$\bullet$}}

\newcommand{\symbolstar}[1][black]{{\color{#1}\ding{72}}}

\definecolor{FGViolet}{rgb}{0.61,0.32,0.61}
\definecolor{FGDarkBlue}{rgb}{0,0,0.6}
\definecolor{FGBlue}{rgb}{0,0,0.8}
\definecolor{FGLightBlue}{rgb}{0.2, 0.6, 0.8}
\definecolor{FGGreen}{rgb}{0.2,0.7,0.2}
\definecolor{FGLightGreen}{rgb}{0.4,1,0.4}
\definecolor{FGYellow}{rgb}{1,0.95,0}
\definecolor{FGOrange}{rgb}{0.95,0.5,0.1}
\definecolor{FGRed}{rgb}{0.8,0,0}
\definecolor{FGWhite}{rgb}{1,1,1}
\definecolor{FGLightGray}{rgb}{0.8,0.8,0.8}
\definecolor{FGGray}{rgb}{0.5,0.5,0.5}
\definecolor{FGDarkGray}{rgb}{0.3,0.3,0.3}
\definecolor{FGBlack}{rgb}{0,0,0}

\begin{document}

\title{Similarity-Transformed Chiral NN+3N Interactions for the Ab Initio Description of \elem{C}{12} and \elem{O}{16}}

\author{Robert Roth}
\email{robert.roth@physik.tu-darmstadt.de}

\author{Joachim Langhammer}
\author{Angelo Calci}
\author{Sven Binder}

\affiliation{Institut f\"ur Kernphysik, Technische Universit\"at Darmstadt, 64289 Darmstadt, Germany}

\author{Petr Navr\'atil}
\affiliation{TRIUMF, 4004 Wesbrook Mall, Vancouver, British Columbia, V6T 2A3 Canada}

\date{\today}

\begin{abstract}  
We present first \emph{ab initio} no-core shell model (NCSM) calculations using similarity renormalization group (SRG) transformed chiral two-nucleon (NN) plus three-nucleon (3N) interactions for nuclei throughout the p-shell, particularly \elem{C}{12} and \elem{O}{16}. By introducing an adaptive importance truncation for the NCSM model space and an efficient $JT$-coupling scheme for the 3N matrix elements, we are able to surpass previous NCSM studies including 3N interactions regarding, both, particle number and model-space size. We present ground and excited states in \elem{C}{12} and \elem{O}{16} for model spaces up to $N_{\max}=12$ including full 3N interactions, which is sufficient to obtain converged results for soft SRG-transformed interactions. We analyze the contributions of induced and initial 3N interactions and probe induced 4N terms through the sensitivity of the energies on the SRG flow parameter. Unlike for light p-shell nuclei, SRG-induced 4N contributions originating from the long-range two-pion terms of the chiral 3N interaction are sizable in \elem{C}{12} and \elem{O}{16}. 

\end{abstract}


\maketitle

%%%%%%%%%%%%%%%%%%%%%%%%%%%%%%%%%%%%%%%%%%%%%%%%%%%%%%%%%%%%%%%%%%%%%%
%%%%%%%%%%%%%%%%%%%%%%%%%%%%%%%%%%%%%%%%%%%%%%%%%%%%%%%%%%%%%%%%%%%%%%
%%%%%%%%%%%%%%%%%%%%%%%%%%%%%%%%%%%%%%%%%%%%%%%%%%%%%%%%%%%%%%%%%%%%%%

Nuclear Hamiltonians constructed within chiral effective field theory (EFT) provide a systematic link between nuclear structure physics and low-energy quantum chromodynamics (QCD). It is a supreme challenge for modern nuclear theory to exploit this link and to apply these interactions consistently in \emph{ab initio} nuclear structure calculations for a wide range of nuclei and observables. This is vital to provide robust QCD-based predictions, e.g. for light exotic nuclei, to constrain approximate nuclear structure approaches, and to understand the relevant QCD mechanisms driving nuclear structure phenomena. At present the most advanced calculations beyond the few-body domain use chiral two-nucleon (NN) interactions at N$^3$LO \cite{EnMa03} and three-nucleon (3N) interactions at N$^2$LO \cite{EpNo02} in the \emph{ab initio} no-core shell model (NCSM) \cite{NaQu09}. A first milestone was the study of the spectroscopy of mid-p-shell nuclei in moderate model spaces using a Lee-Suzuki (LS) transformed Hamiltonian \cite{NaGu07}, which proved the predictive power of chiral Hamiltonians and the need to include the 3N interaction consistently. Recently, light p-shell nuclei were studied in the NCSM employing a consistent similarity renormalization group (SRG) transformation of the chiral NN+3N Hamiltonian \cite{JuNa09,JuNa11}. The SRG transformation provides a model-space independent Hamiltonian with superior convergence properties that can be used universally in a variety of many-body approaches \cite{RoNe10,BoFu10}.

In this Letter we present the first \emph{ab initio} calculations of nuclei throughout the whole p-shell including \elem{C}{12} and \elem{O}{16} using SRG-transformed chiral NN+3N interactions. Through a combination of conceptual and computational developments we are able to extend the range of previous NCSM studies using full 3N interactions to significantly larger model spaces and particle numbers.

%%%%%%%%%%%%%%%%%%%%%%%%%%%%%%%%%%%%%%%%%%%%%%%%%%%%%%%%%%%%%%%%%%%%%%
%%%%%%%%%%%%%%%%%%%%%%%%%%%%%%%%%%%%%%%%%%%%%%%%%%%%%%%%%%%%%%%%%%%%%%
%%%%%%%%%%%%%%%%%%%%%%%%%%%%%%%%%%%%%%%%%%%%%%%%%%%%%%%%%%%%%%%%%%%%%%
\paragraph{SRG-Transformed NN+3N Interactions.}

A crucial step for NCSM calculations beyond the lightest isotopes is the unitary transformation of the initial Hamiltonian in order to improve the convergence behavior with respect to the size of the many-body model space. In addition to the LS similarity transformation, which is tailored to decouple the NCSM model space from the excluded space, several model-space independent unitary transformations, e.g. the unitary correlation operator method (UCOM) and the SRG, have been introduced \cite{BoFu10,RoNe10}. Here, we focus on the SRG, mainly because of its simplicity and flexibility. 

In the SRG framework the unitary transformation of an operator, e.g. the Hamiltonian, is formulated in terms of a flow equation
$\frac{d}{d\alpha} \op{H}_\alpha = \comm{\op{\eta}_{\alpha}}{\op{H}_\alpha}$
with a continuous flow parameter $\alpha$. The initial condition for the solution of this flow equation is given by the `bare' chiral Hamiltonian. The physics of the SRG evolution is governed by the anti-hermitian generator $\op{\eta}_{\alpha}$. A specific form widely used in nuclear physics \cite{BoFu10,RoNe10} is given by $\op{\eta}_\alpha = m_N^2\; \comm{\op{T}_{\text{int}}}{\op{H}_\alpha}$, where $m_N$ is the nucleon mass and $\op{T}_{\text{int}}=\op{T}-\op{T}_{\text{cm}}$ is the intrinsic kinetic energy operator.  This generator drives the Hamiltonian towards a diagonal form in a basis of eigenstates of the intrinsic kinetic energy. 

Along with the pre-diagonalization of the Hamiltonian, which is the reason for the transformation in the first place, the SRG induces many-body operators beyond the particle-rank of the initial Hamiltonian. Only if all the induced terms up to the $A$-body level are kept, the transformation is unitary and the spectrum of the Hamiltonian in an exact $A$-body calculation is unchanged and independent of the flow parameter $\alpha$. In practice we have to truncate the evolution at a particle rank $n<A$, thus violating formal unitarity. In this situation we can use the flow-parameter $\alpha$ as a diagnostic tool to quantify the contribution of omitted beyond-$n$-body terms.

Whereas the SRG transformation at two-body level has been used for some time \cite{BoFu07b,RoRe08,RoNe10}, the solution of the evolution equations at three-body level was demonstrated only recently \cite{JuNa09,JuNa11}. In view of the application in the NCSM it is convenient to solve the flow equation for the three-body system using a harmonic-oscillator (HO) Jacobi-coordinate basis \cite{NaKa00}. The intermediate sums in the 3N Jacobi basis are truncated at $N_{\max}=40$ for channels with $J\leq5/2$ and ramp down linearly to $N_{\max}=24$ for $J\geq13/2$. Based on this and the corresponding solution of the flow equation in two-body space (using either a partial-wave momentum or harmonic-oscillator representation) we extract the irreducible two- and three-body terms of the Hamiltonian for the use in $A$-body calculations. 
  
We have made major technical improvements regarding the SRG transformation, reducing the computational effort by three orders of magnitude compared to Ref. \cite{JuNa11}, e.g., by using a solver with adaptive step-size and optimized matrix operations. Furthermore, we have developed a transformation from 3N Jacobi matrix elements to a $JT$-coupled representation with a highly efficient storage scheme, which allows us to handle 3N matrix-element sets of unprecedented size. A detailed discussion of these aspects is presented elsewhere.

%%%%%%%%%%%%%%%%%%%%%%%%%%%%%%%%%%%%%%%%%%%%%%%%%%%%%%%%%%%%%%%%%%%%%%
%%%%%%%%%%%%%%%%%%%%%%%%%%%%%%%%%%%%%%%%%%%%%%%%%%%%%%%%%%%%%%%%%%%%%%
%%%%%%%%%%%%%%%%%%%%%%%%%%%%%%%%%%%%%%%%%%%%%%%%%%%%%%%%%%%%%%%%%%%%%%
\paragraph{Importance-Truncated NCSM.}

Based on the SRG-evolved Hamiltonian we treat the many-body problem in the NCSM, i.e., we solve the large-scale eigenvalue problem of the Hamiltonian, represented in a many-body basis of HO Slater determinants truncated w.r.t. the maximum HO excitation energy $N_{\max}\hbar\Omega$. In order to cope with the factorial growth of the basis dimension with $N_{\max}$ and particle number $A$, we use the importance-truncation (IT) scheme introduced in Refs.~\cite{RoNa07,Roth09}. 
The IT-NCSM uses an importance measure $\kappa_{\nu}$ for the individual basis states $\ket{\Phi_{\nu}}$ derived from many-body perturbation theory and retains only states with $|\kappa_\nu|$ above a threshold $\kappa_{\min}$ in the model space. Through a variation of the threshold and an \emph{a posteriori} extrapolation $\kappa_{\min}\to0$ the contribution of discarded states is recovered. We use the sequential update scheme discussed in Ref.~\cite{Roth09}, which connects to the full NCSM model space and thus the exact NCSM results in the limit of vanishing threshold. In the following we always report threshold-extrapolated results including an estimate for the extrapolation uncertainties. For the present application we have extended the IT-NCSM to include full 3N interactions. Using the $JT$-coupled 3N matrix elements we are able to perform calculations up to $N_{\max}=12$ or $14$ for all p-shell nuclei with moderate computational resources. Due to the $JT$-coupling, we can keep all 3N matrix elements in memory using a fast on-the-fly decoupling.

%%%%%%%%%%%%%%%%%%%%%%%%%%%%%%%%%%%%%%%%%%%%%%%%%%%%%%%%%%%%%%%%%%%%%%
%%%%%%%%%%%%%%%%%%%%%%%%%%%%%%%%%%%%%%%%%%%%%%%%%%%%%%%%%%%%%%%%%%%%%%
%%%%%%%%%%%%%%%%%%%%%%%%%%%%%%%%%%%%%%%%%%%%%%%%%%%%%%%%%%%%%%%%%%%%%%
\paragraph{Ground-State Energies.}

We first focus on IT-NCSM calculations for the ground states of \elem{He}{4}, \elem{Li}{6}, \elem{C}{12}, and \elem{O}{16} using SRG-transformed chiral NN+3N interactions. Throughout this work we use the chiral NN interaction at N$^3$LO of Entem and Machleidt \cite{EnMa03} and the 3N interaction at N$^2$LO \cite{Navr07} with low energy constants determined from the triton binding energy and $\beta$-decay half-live \cite{GaQu09}. In order to disentangle the effects of the initial and the SRG-induced 3N contributions, we consider three different Hamiltonians.
\emph{(1) NN only}: starting from the chiral NN interaction only the SRG-evolved NN contributions are kept.
\emph{(2) NN+3N-induced}: starting from the chiral NN interaction the SRG-evolved NN and the induced 3N terms are kept.
\emph{(3) NN+3N-full}: starting from the chiral NN+3N interaction the SRG-evolved NN and all 3N terms are kept. 
For each Hamiltonian we assess the dependence of the observables, here the ground-state energies, on the flow-parameter $\alpha$. We use the five values $\alpha=0.04\,\text{fm}^4$, $0.05\,\text{fm}^4$, $0.0625\,\text{fm}^4$, $0.08\,\text{fm}^4$, and $0.16\,\text{fm}^4$, which correspond to momentum scales $\Lambda=\alpha^{-1/4}=2.24\,\text{fm}^{-1}$, $2.11\,\text{fm}^{-1}$, $2\,\text{fm}^{-1}$, $1.88\,\text{fm}^{-1}$, and $1.58\,\text{fm}^{-1}$, respectively. For extrapolations to infinite model space, $N_{\max}\to\infty$, we use simple exponential fits based on the last 3 or 4 data points. The extrapolated energy is given by the average of the two extrapolations, the uncertainty by the difference.

%%%%%%%%%%%%%%%%%%%%%%%%%%%%%%%%%%%%%%%%%%%%%%%%%%%%%%%%%%%%%%%%%%%%
\begin{figure}
\includegraphics[width=1\columnwidth]{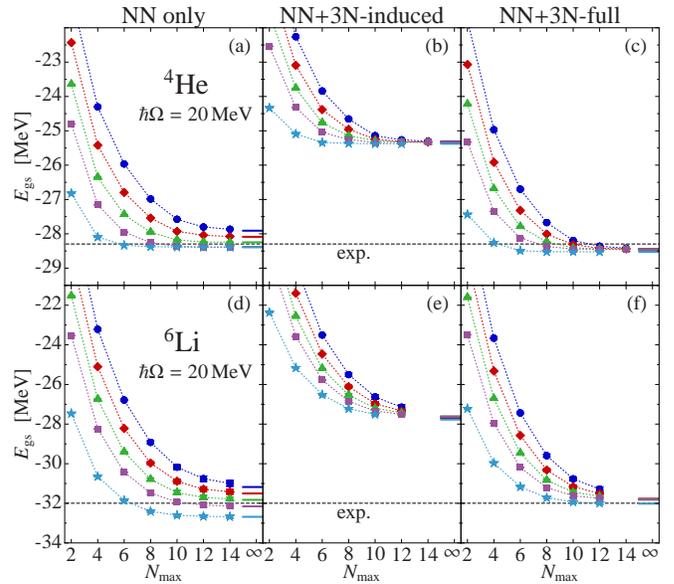}
\caption{(color online) IT-NCSM ground-state energies for \elem{He}{4} and \elem{Li}{6} as function of $N_{\max}$ for the three types of Hamiltonians (see column headings) for a range of flow parameters: $\alpha=0.04\,\text{fm}^4$ (\symbolcircle[FGBlue]), $0.05\,\text{fm}^4$ (\symboldiamond[FGRed]), $0.0625\,\text{fm}^4$ (\symboltriangle[FGGreen]), $0.08\,\text{fm}^4$ (\symbolbox[FGViolet]), and $0.16\,\text{fm}^4$ (\symbolstar[FGLightBlue]). Error bars indicate the uncertainties of the threshold extrapolations. The bars at the right-hand-side of each panel indicate the results of exponential extrapolations of the individual $N_{\max}$-sequences (see text).}
\label{fig:itncsm_He4Li6}
\end{figure}
%%%%%%%%%%%%%%%%%%%%%%%%%%%%%%%%%%%%%%%%%%%%%%%%%%%%%%%%%%%%%%%%%%%%

The ground-state energies obtained in IT-NCSM calculations for \elem{He}{4} and \elem{Li}{6} with the three Hamiltonians are summarized in Fig.~\ref{fig:itncsm_He4Li6}. Analogous calculations in the full NCSM for the same SRG-evolved initial Hamiltonian have been presented in Ref.~\cite{JuNa09} for \elem{He}{4} and in Ref.~\cite{JuNa11} for \elem{Li}{6}. We have cross-checked our results with Refs.~\cite{JuNa09,JuNa11} and found excellent agreement. 

The first and foremost effect of the SRG transformation is the acceleration of the convergence of NCSM calculations with $N_{\max}$. With increasing $\alpha$ the convergence is systematically improved for all three versions of the Hamiltonian. With the initial Hamiltonian, i.e. $\alpha=0$, even the large model spaces we use here are not sufficient to even obtain meaningful extrapolations, with the exception of the tightly-bound $^4$He.

For the NN-only Hamiltonian Fig.~\ref{fig:itncsm_He4Li6} shows a clear $\alpha$-dependence of the extrapolated ground-state energies for \elem{He}{4} and \elem{Li}{6}, hinting at sizable SRG-induced 3N contributions. When including those induced 3N terms, i.e. when using the NN+3N-induced Hamiltonian, the extrapolated ground-state energies are shifted significantly and become $\alpha$-independent within the uncertainties of the $N_{\max}$-extrapolation. Thus, induced contributions beyond the 3N level originating from the initial NN interaction are negligible in the $\alpha$-range considered here, indicating that the NN+3N-induced Hamiltonian is unitarily equivalent to the initial NN Hamiltonian. The extrapolated ground-state energies for different $\alpha$ are summarized in Tab.~\ref{tab:itncsm}. 

%%%%%%%%%%%%%%%%%%%%%%%%%%%%%%%%%%%%%%%%%%%%%%%%%%%%%%%%%%%%%%%%%%%%%%
\begin{table}[b]
\caption{Summary of $N_{\max}$-extrapolated IT-NCSM ground-state energies in MeV for $\hbar\Omega=20\,\text{MeV}$ (see text).}
\label{tab:itncsm}
\begin{ruledtabular}
\begin{tabular}{l l l l l l}
  & \multicolumn{1}{c}{$\alpha\;[\text{fm}^4]$} & \multicolumn{1}{c}{\elem{He}{4}} & \multicolumn{1}{c}{\elem{Li}{6}} & 
    \multicolumn{1}{c}{\elem{C}{12}} & \multicolumn{1}{c}{\elem{O}{16}} \\
\hline 
NN            & 0.05   & -28.08(2)  & -31.5(2)  &  -99.1(6)  & -161.0(2) \\
only          & 0.0625 & -28.25(1)  & -31.8(1)  & -101.4(3)  & -164.9(6) \\
              & 0.08   & -28.38(1)  & -32.2(1)  & -103.7(2)  & -170.2(4) \\
\hline
NN+           & 0.05   & -25.33(1)  & -27.7(2)  &  -76.9(2)  & -119.5(3) \\
3N-ind.       & 0.0625 & -25.34(1)  & -27.6(2)  &  -77.2(1)  & -119.7(6) \\
              & 0.08   & -25.34(1)  & -27.6(1)  &  -77.4(2)  & -119.5(2) \\
\hline
NN+           & 0.05   & -28.45(3)  & -31.8(2)  &  -96.1(4)  & -143.7(2) \\
3N-full       & 0.0625 & -28.45(1)  & -31.8(1)  &  -96.8(3)  & -145.6(2) \\
              & 0.08   & -28.46(1)  & -31.8(1)  &  -97.6(1)  & -147.8(1) \\
\hline
exp.          &        & -28.30     & -31.99    &  -92.16    & -127.62   \\
\end{tabular}
\end{ruledtabular}
\end{table}
%%%%%%%%%%%%%%%%%%%%%%%%%%%%%%%%%%%%%%%%%%%%%%%%%%%%%%%%%%%%%%%%%%%%%%%%
 
By including the initial chiral 3N interaction, i.e., by using the NN+3N-full Hamiltonian, the ground-state energies are lowered and are in good agreement with experiment for both, \elem{He}{4} and \elem{Li}{6}. As for the NN+3N-induced there is no sizable $\alpha$-dependence in the range considered here. We conclude that induced 3N terms originating from the initial NN interaction are important, but that induced 4N (and higher) terms are not relevant for light p-shell nuclei, since the ground-state energies obtained with the NN+3N-induced and the NN+3N-full Hamiltonian are practically $\alpha$-independent.

%%%%%%%%%%%%%%%%%%%%%%%%%%%%%%%%%%%%%%%%%%%%%%%%%%%%%%%%%%%%%%%%%%%%
\begin{figure}
\includegraphics[width=1\columnwidth]{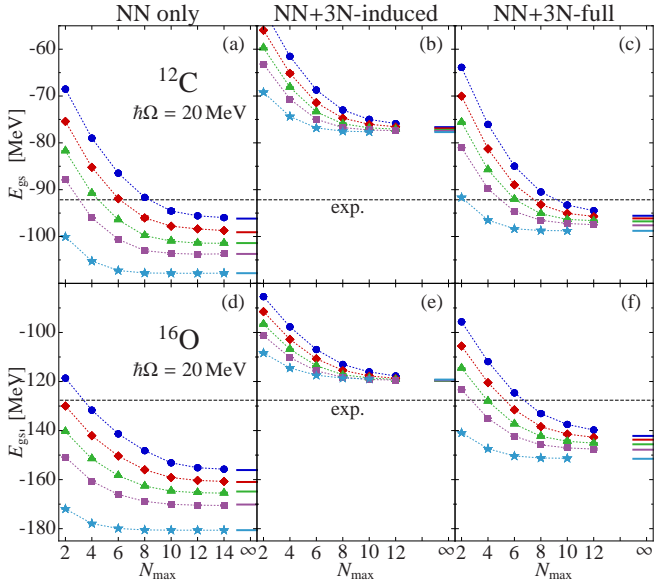}
\caption{(color online) IT-NCSM ground-state energies for \elem{C}{12} and \elem{O}{16} as function of $N_{\max}$ for the three types of Hamiltonians and a range of flow parameters (for details see Fig. \ref{fig:itncsm_He4Li6}).}
\label{fig:itncsm_C12O16}
\end{figure}
%%%%%%%%%%%%%%%%%%%%%%%%%%%%%%%%%%%%%%%%%%%%%%%%%%%%%%%%%%%%%%%%%%%%

This picture changes if we consider nuclei in the upper p-shell. In Fig.~\ref{fig:itncsm_C12O16} we show the first accurate \emph{ab initio} calculations for the ground states of \elem{C}{12} and \elem{O}{16} starting from chiral NN+3N interactions. By combining the IT-NCSM with the $JT$-coupled storage scheme for the 3N matrix elements we are able to reach model spaces up to $N_{\max}=12$ for the upper p-shell at moderate computational cost. Previously, even the most extensive NCSM calculations including full 3N interactions were limited to $N_{\max}=8$ in this regime \cite{MaVa11}. As evident from the $N_{\max}$-dependence of the ground-state energies, this increase in $N_{\max}$ is vital for obtaining precise extrapolations.

The general pattern for \elem{C}{12} and \elem{O}{16} is similar to the light p-shell nuclei: The NN-only Hamiltonian exhibits a severe $\alpha$-dependence indicating sizable induced 3N contributions. Their inclusion in the NN+3N-induced Hamiltonian leads to ground-state energies that are practically independent of $\alpha$, confirming that induced 4N contributions are irrelevant when starting from the NN interaction only. Therefore, the NN+3N-induced results can be considered equivalent to a solution  for the initial NN interaction. The \elem{O}{16} binding energy per nucleon of $7.48(4)\,\text{MeV}$ is in good agreement with a recent coupled-cluster $\Lambda$-CCSD(T) result of $7.56\,\text{MeV}$ for the `bare' chiral NN interaction \cite{HaPa10}. 

In contrast to light nuclei the ground-state energies of \elem{C}{12} and \elem{O}{16} obtained with the NN+3N-full Hamiltonian do show a significant $\alpha$-dependence, as evident from Fig.~\ref{fig:itncsm_C12O16}(c) and (f). The inclusion of the initial chiral 3N interaction does induce 4N contributions whose omission leads to the $\alpha$-dependence. 

%%%%%%%%%%%%%%%%%%%%%%%%%%%%%%%%%%%%%%%%%%%%%%%%%%%%%%%%%%%%%%%%%%%%
\begin{figure}
\includegraphics[width=1\columnwidth]{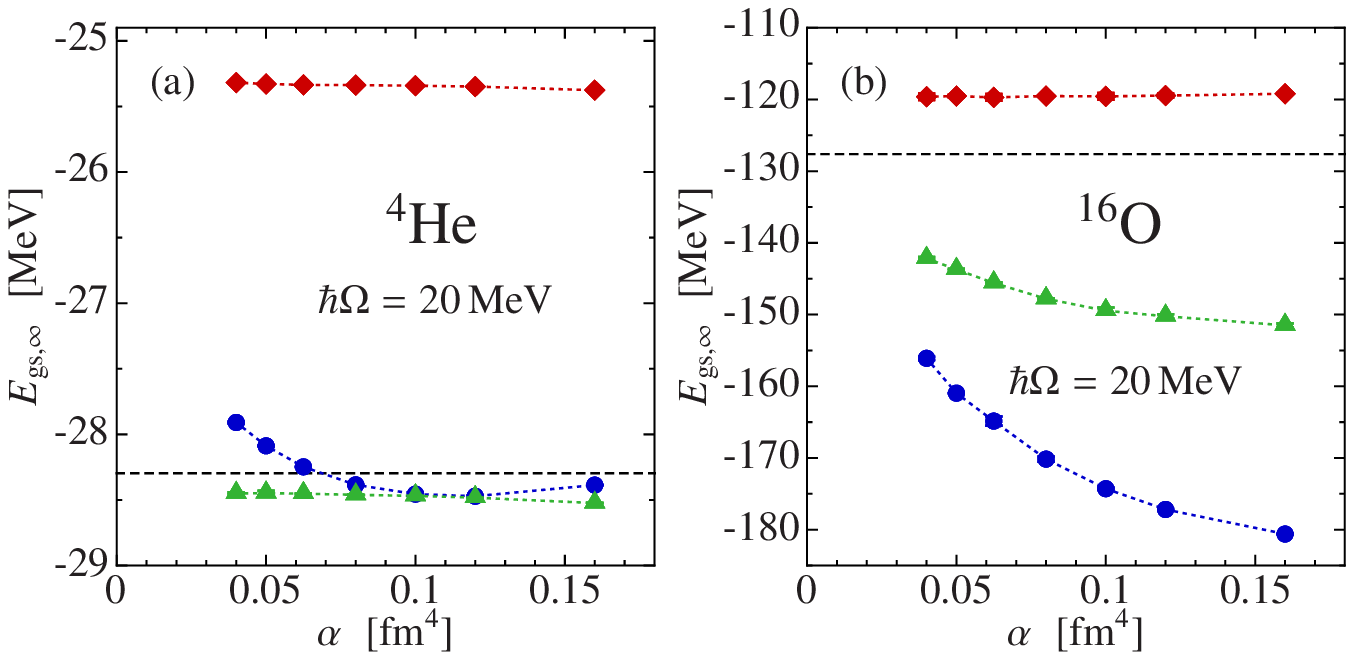}
\caption{(color online) $N_{\max}$-extrapolated ground-state energies of \elem{He}{4} and \elem{O}{16} as function of the flow parameter $\alpha$ for the NN-only (\symbolcircle[FGBlue]), the NN+3N-induced (\symboldiamond[FGRed]), and the NN+3N-full Hamiltonian (\symboltriangle[FGGreen]).}
\label{fig:itncsm_He4O16_alphaDep}
\end{figure}
%%%%%%%%%%%%%%%%%%%%%%%%%%%%%%%%%%%%%%%%%%%%%%%%%%%%%%%%%%%%%%%%%%%%

A direct comparison of the $\alpha$-dependence of the extrapolated ground-state energies for \elem{He}{4} and \elem{O}{16} is presented in Fig.~\ref{fig:itncsm_He4O16_alphaDep}. For both nuclei, the NN-only Hamiltonian exhibits a sizable variation of the ground-state energies of about 25 MeV (0.7 MeV) for \elem{O}{16} (\elem{He}{4}) in the range from $\alpha=0.04\,\text{fm}^4$ to $0.16\,\text{fm}^4$. The inclusion of the induced 3N terms eliminates this $\alpha$-dependence. The inclusion of the initial 3N interaction again generates an $\alpha$-dependence of about 10 MeV for \elem{O}{16}. Note that the induced 4N (and higher) contributions that are needed to compensate the $\alpha$-dependence for \elem{O}{16} reach about half the size of the total 3N contribution in the SRG-transformed Hamiltonian. This is evidence that the hierarchy of the many-body forces in chiral EFT may not be preserved by the SRG transformation.

We have analyzed the role of the different terms of the chiral 3N interaction at N$^2$LO regarding the induced 4N terms. By setting the low-energy constants associated with either the three-body contact ($c_E$), the one-pion term ($c_D$), or the long-range two-pion terms ($c_{1,3,4}$) to zero and readjusting the remaining $c_D$ or $c_E$ to reproduce the triton binding energy, 
we find a strong reduction of the $\alpha$-dependence for the $c_{1,3,4}=0$ interaction. Thus, the long-range two-pion exchange contribution to  the chiral 3N drives the induced 4N contributions.

%%%%%%%%%%%%%%%%%%%%%%%%%%%%%%%%%%%%%%%%%%%%%%%%%%%%%%%%%%%%%%%%%%%%%%
%%%%%%%%%%%%%%%%%%%%%%%%%%%%%%%%%%%%%%%%%%%%%%%%%%%%%%%%%%%%%%%%%%%%%%
%%%%%%%%%%%%%%%%%%%%%%%%%%%%%%%%%%%%%%%%%%%%%%%%%%%%%%%%%%%%%%%%%%%%%%
\paragraph{Excitation Spectra.}

In addition to ground-state energies the IT-NCSM provides full access to spectra and all spectroscopic observables. As an example we consider the effect of 3N terms on the excitation spectrum of \elem{C}{12}. The $\alpha$-dependence of the excitation energies follows a similar pattern as the ground-state energies but is much weaker. The variation of the excitation energy of the first excited $2^+$ state across the $\alpha$-range discussed before is on the order of a few 100 keV for the NN+3N-full Hamiltonian. Thus, meaningful calculations of excitation spectra are already possible at the NN+3N-full level without accounting for the induced 4N contributions.

%%%%%%%%%%%%%%%%%%%%%%%%%%%%%%%%%%%%%%%%%%%%%%%%%%%%%%%%%%%%%%%%%%%%
\begin{figure}
\includegraphics[width=1\columnwidth]{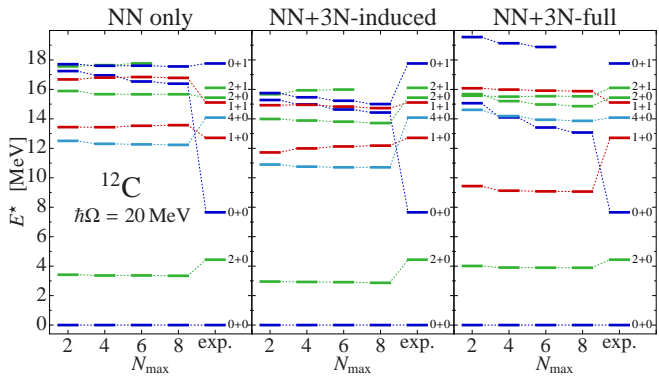}
\caption{(color online) Excitation spectrum for the lowest positive-parity states (labelled $J\pi T$) in \elem{C}{12} for the NN-only, the NN+3N-induced, and the NN+3N-full Hamiltonian with $\alpha=0.08\,\text{fm}^4$.}
\label{fig:itncsm_C12_spectrum}
\end{figure}
%%%%%%%%%%%%%%%%%%%%%%%%%%%%%%%%%%%%%%%%%%%%%%%%%%%%%%%%%%%%%%%%%%%%

To illustrate the sensitivity of the excitation spectrum on the induced and initial 3N interactions we present in Fig.~\ref{fig:itncsm_C12_spectrum} the first six excited states of positive parity for fixed $\alpha=0.08\,\text{fm}^4$. Overall, the spectrum agrees well with the calculation presented in \cite{NaGu07} for $N_{\max}=6$ using a different LS-transformed chiral 3N interaction. Considering the changes in the spectrum caused by the inclusion of the induced 3N interaction (change from NN-only to NN+3N-induced) and the initial 3N interaction (change from NN+3N-induced to NN+3N-full), it is evident that the induced 3N terms only lead to an over-all compression of the spectrum, whereas the initial 3N interaction affects the individual levels in very different ways, thus changing the level ordering. In comparison to the experimental spectrum, the agreement for the first $2^+$ and $4^+$ states are significantly improved by the inclusion of the initial 3N interaction. The first $1^+$ state, however, is pushed too low by the initial 3N interaction, ending up in clear disagreement with experiment. The only other state which is not well described is the first excited $0^+$ state, the famous Hoyle state. Though the 3N interaction improves the situation, the NCSM basis is not well suited for the description of strong $\alpha$-clustering in this state, which gives rise to the very slow convergence (for a complementary approach based on chiral EFT see \cite{EpKr11}).

%%%%%%%%%%%%%%%%%%%%%%%%%%%%%%%%%%%%%%%%%%%%%%%%%%%%%%%%%%%%%%%%%%%%%%
%%%%%%%%%%%%%%%%%%%%%%%%%%%%%%%%%%%%%%%%%%%%%%%%%%%%%%%%%%%%%%%%%%%%%%
%%%%%%%%%%%%%%%%%%%%%%%%%%%%%%%%%%%%%%%%%%%%%%%%%%%%%%%%%%%%%%%%%%%%%%
\paragraph{Conclusions.}

We have presented the first \emph{ab initio} nuclear structure calculations for nuclei in the upper p-shell based on SRG-evolved chiral NN+3N interactions. By introducing $JT$-coupled 3N matrix elements and the importance truncation of the NCSM model space we are able to surpass all previous NCSM studies including 3N interactions regarding particle number and model-space size. We find that, unlike for light p-shell nuclei, SRG-induced 4N contributions originating from the long-range two-pion terms of the chiral 3N interaction are sizable in \elem{C}{12} and \elem{O}{16}. Based on this finding alternative SRG generators can be explored with the aim to suppress induced 4N terms. Together with the advances presented here, this will enable detailed \emph{ab initio} studies of p- and sd-shell nuclei starting from full chiral NN+3N interactions.

%%%%%%%%%%%%%%%%%%%%%%%%%%%%%%%%%%%%%%%%%%%%%%%%%%%%%%%%%%%%%%%%%%%%%%
%%%%%%%%%%%%%%%%%%%%%%%%%%%%%%%%%%%%%%%%%%%%%%%%%%%%%%%%%%%%%%%%%%%%%%
%%%%%%%%%%%%%%%%%%%%%%%%%%%%%%%%%%%%%%%%%%%%%%%%%%%%%%%%%%%%%%%%%%%%%%
\paragraph{Acknowledgments.}

Numerical calculations have been performed at the J\"ulich Supercomputing Centre and at LOEWE-CSC. Supported by the DFG through contract SFB 634 and by the Helmholtz International Center for FAIR (HIC for FAIR). P. N. acknowledges partial support from the UNEDF SciDAC Collaboration DOE Grant DE-FC02-07ER41457 and the NSERC grant No. 401945-2011.

%%%%%%%%%%%%%%%%%%%%%%%%%%%%%%%%%%%%%%%%%%%%%%%%%%%%%%%%%%%%%%%%%%%%%%
%%%%%%%%%%%%%%%%%%%%%%%%%%%%%%%%%%%%%%%%%%%%%%%%%%%%%%%%%%%%%%%%%%%%%%
%%%%%%%%%%%%%%%%%%%%%%%%%%%%%%%%%%%%%%%%%%%%%%%%%%%%%%%%%%%%%%%%%%%%%%
%\bibliography{/u/rroth/Papers/bib/bib_nucl}

%%%%%%%%%%%%%%%%%%%%%%%%%%%%%%%%%%%%%%%%%%%%%%%%%%%%%%%%%%%%%%%%%%%%%%
%%%%%%%%%%%%%%%%%%%%%%%%%%%%%%%%%%%%%%%%%%%%%%%%%%%%%%%%%%%%%%%%%%%%%%
%%%%%%%%%%%%%%%%%%%%%%%%%%%%%%%%%%%%%%%%%%%%%%%%%%%%%%%%%%%%%%%%%%%%%%
\end{document}